Working Paper
TPRC Conference on Information, Communications, and Internet Policy
October 2001

**Is the Commercial Mass Media Necessary, or Even Desirable, for Liberal Democracy?**

Neil Netanel[*]

As James Madison aptly put it: "A popular Government, without popular information, or the means of acquiring it, is but a Prologue to a Farce or a Tragedy; or, perhaps both."[1] In our day, certainly in the United States and to a growing extent in other developed democracies, most citizens most of the time acquire popular information from the commercial mass media. It hasn't always been that way and, perhaps, needn't be that way in the future. In this essay, I ask whether the commercial media is the best vehicle for gathering and distributing information and opinion to citizens of advanced liberal democracies.

The role of commercial media in liberal democracy has been the subject of renewed controversy in light of peer-to-to peer alternatives made possible by the Internet and other digital communications technology. I, and other commentators, have asserted that, despite all their imperfections, the commercial media play and will continue to play an important "Fourth Estate" role in liberal democratic governance. Media critics have countered that (1) liberal democratic governance would be better off if peer-to-peer communications were to assume the primary role in the dissemination of information and opinion and (2) digital communications makes that transition possible.

This paper seeks to push the ball forward on this issue. Part I broadly outlines the principal tenets of liberal democracy. Part II describes the commercial media's purported "Fourth Estate" role and briefly notes the ways it falls short of realizing it. Parts III and IV limn two alternatives to the commercial media: political-party financed press and government-funded media. These Parts focus on the advantages and disadvantages of those alternatives as compared to the commercial media. Part V presents the proffered alternative of peer-to-peer communication. Part VI critiques that alternative.

---

[*] Arnold, White & Durkee Centennial Professor of Law, University of Texas School of Law.
[1] Letter from James Madison to W.T. Barry (Aug. 4, 1822), reprinted in 9 THE WRITINGS OF JAMES MADISON 103, 103 (Gaillard Hunt ed., 1910).

I. Liberal Democracy

"Liberal democracy" has multiple meanings and iterations. One's understanding of "liberal democracy" can heavily color one's view and expectations concerning the desired role of the commercial media and its alternatives.[2] Accordingly, I briefly outline my understanding of "liberal democracy." In so doing, I build upon Ed Baker's model of "complex democracy," as distinct from elitist, interest group pluralist, and republican democracy.[3] This model, I believe, captures the most salient common characteristics and aspirations of advanced, developed-country democracies at the beginning of the twenty-first century.

Liberal democracy presents numerous fault lines. These include tensions between individual autonomy and democratic process and between interest-group pluralism and more republican notions of a common public good. Contrary to the view of some theorists, liberal democratic theory and practice incorporates and seeks to accommodate all of those elements.[4] It includes values and legal doctrine that highlight individual autonomy, interest group pluralism, and common public discourse. As Ed Baker puts it, contrasting his model of "complex democracy" with interest group pluralism, civic republicanism, and other unidimensional variants: "A more 'realistic' theory would assume that a participatory democracy would and should encompass arenas where both individuals and groups look for and create common ground, that is, common goods, but where they also advance their own individual and group values and interests. Moreover, normatively, it is difficult to argue that either type of political striving is inappropriate for an ethical person or within a justifiable politics."[5]

---

[2] Ed Baker has convincingly detailed this point with respect to "democracy" and the media. *See* C. Edwin Baker, *The Media That Citizens Need*, 147 U. PENN. L. REV. 317 (1998).
[3] *See id*. at 335-43.
[4] For a cogent refutation of the notion that rights-based liberalism is inherently hostile to democracy, *see* STEPHEN HOLMES, PASSIONS AND CONSTRAINT: ON THE THEORY OF LIBERAL DEMOCRACY 29-36 (1995). In addition, as Ed Rubin insightfully argues, the often overblown rhetoric of "rights" and "democracy" fails to capture the essence of most government decision making in a modern "liberal democratic" administrative state. *See* Edward L. Rubin, *Getting Past Democracy*, 149 U. PA. L. REV. 711 (2001).
[5] Baker, *supra* note 2, at 336.

Significantly, and this is certainly the experience of liberal democratic nations, complex democracy necessarily encompass multiple concentrations of power. Interest groups, civic associations, business firms, political parties, labor unions, and government subdivisions and agencies vie and compete for power and influence in a wide variety of arenas.[6] The dispersion of power and consequent multiple concentrations of power are both inevitable to and part of the fabric of a pluralist, liberal democracy. They are part and parcel of divided and limited government and of individual's right and ability to organize in common social, political, and commercial enterprise.

As I will discuss below, in designing a policy regarding the dissemination of information and expression in a liberal democratic society, we must take account of both the various values that comprise "complex democracy" and the fact of multiple concentrations of political and economic power. An information policy for liberal democracy "properly includes a discursive search for common goods and agreement, but it also must involve recognition of goods that are not common and about which there must be fair bargaining and compromise."[7]

## II.  The Commercial Media's "Fourth Estate" Role

Echoing a widely held view of the role of the institutional press, I have argued elsewhere that "liberal democracy requires an institutional media that possesses the financial strength to reach a mass audience and engage in sustained investigative reporting, free from a potentially corrupting dependence on state subsidy."[8] Underlying that view is an understanding of the role that the commercial media plays in supporting liberal democracy. In acting as the "Fourth Estate," the press is said to fulfill three basic functions. These are watchdog, public discourse facilitation, and trustworthy supplier of information.

I use those three functions as the framework for our analysis. I posit that liberal democracy as I have described it requires some institution or institutions that serve those functions. Accordingly, I use the extent to which the commercial media and its alternatives actually serve the functions as a vehicle for testing their relative desirability for liberal democracy. More particularly, I take the status quo– a system of free expression dominated by the commercial media – as my starting point. I take cognizance of the abundant criticism of the commercial media and agree that the media imperfectly serves as watchdog, public discourse facilitation, and trustworthy supplier of information (though we may disagree on the extent to which the commercial media falls short of the ideal). But I then look at some leading alternatives, particular peer-to-peer communication -- and ask whether they have performed or are likely to perform any better.

---

[6] *See* notes 73 - 74 *infra*.
[7] Baker, *supra* note 2, at 339.
[8] Neil Weinstock Netanel, *Market Hierarchy and Copyright in Our System of Free Expression*, 53 VANDERBILT L. REV. 1879, 1919 (2000).

I begin by outlining the three functions. I first describe in their ideal form, the way they are supposed to work. I then briefly rehearse some of the ways in which the commercial media fall short of this ideal.

*A. Watchdog*

To the extent it fulfills the watchdog function, the commercial media catalyzes and, to a degree, embodies public opinion in the face of both government authority and private centers of power. It serves to check the power -- and abuse of power -- of government, business, political parties, interest groups, and associations.

The media acts as watchdog, first, by providing the public with information that the public can then use to judge the politically and economically powerful. Significantly in that regard, the media engages in investigative reporting.[9] It aggressively seeks out information and scrutinizes those in positions of power, rather than merely acting as a passive conduit for relaying the information given to it by the powerful. Armed with that information, the public can then vote out elected officials whom the press exposes as incompetent, corrupt, or out-of-touch with majority views. Citizens can also boycott or otherwise organize against businesses or other associations that the press exposes as unworthy (as those citizens would define it). A prime, recent example is the reaction of the public and elected officials to media reporting of tobacco company executives' perjurous denials that they long knew of nicotine's addictive properties.[10]

---

[9] *See, generally*, DAVID L. PROTESS ET. AL., THE JOURNALISM OF OUTRAGE: INVESTIGATIVE REPORTING AND AGENDA BUILDING IN AMERICA (1991).

In addition, the media fulfills its watchdog role by, in some sense, representing the public before public and private officials. Officials react to media reporting with the assumption that it reflects public opinion.[11] They also alter their conduct out of concern for the possibility of that conduct coming under press scrutiny. Thus, the mere presence of the media helps check public and private power. At times, of course, officials simply try to cover-up illicit activity. But overall the ever-present glare of the media gives officials a strong incentive to desist from abuse of power.

### B. Public Discourse Facilitation

In addition to their watchdog function, media enterprises help to lay a foundation for public discourse. Liberal democratic governance requires some measure of *public* discourse, some means for identifying issues of widespread concern and some forum for confronting opposing perspectives.[12] Individuals may have widely divergent interests and preferences. But democratic politics requires at least some shared understanding about what are the most significant issues facing the polity as a whole. And in that regard, studies show that the public can focus only on a half a dozen or so issues at any given time.[13] As a result, politics is not only about prevailing on contentious issues. It is, just as importantly, a struggle to place one's favored issues on the public agenda.

The commercial media serve as conduits for discursive exchange. But, of course, they are not mere passive common carriers. Rather, they present a forum for mediated and edited deliberation and debate. In so doing, the media both delimits a range of

---

[10] See Netanel, *supra* note 8, at 1923-25 (discussing 60 Minutes interview with tobacco company whistle blower Jeffrey Wigand).

[11] *See* PROTESS ET. AL., *supra* note 9, at 244-49 (noting, on the basis of detailed case studies of investigative reporting, that government officials tend to respond to investigative reporters and media exposés before interest groups or the public take up the issue, treating the press as if it were the public).

[12.] See CASS R. SUNSTEIN, FREE MARKETS AND SOCIAL JUSTICE 186-87 (1997) (contending that liberal democracy requires realm of discursive exchange in which citizens can test their preferences and produce better collective decisions).

[13] See Maxwell McCombs, et. al., *Issues in the News and the Public Agenda: The Agenda-Setting Tradition, in* PUBLIC OPINION AND THE COMMUNICATION OF CONSENT 281, 292 (Theodore L. Glasser &

passable opinion and actively contributes to shaping a rough consensus regarding what are the important public issues that need to be addressed.[14] Such media agenda-setting has positive as well as negative ramifications. In so doing, the mass media narrows the public agenda to a limited set of resonant issues, which is in some sense antithetical to expressive diversity. But on the other hand, public debate in a highly pluralistic, advanced democratic state cannot proceed without some measure of broad public consensus on the major priorities of the day.

Agenda-setting also has value for individual self-realization. We define ourselves within and against social groups. For that reason, we generally want to read, see, and hear at least in part what we think others of our social group are reading, seeing, and hearing.[15] We want to experience cultural events and phenomena jointly with others and to share a common basis for conversation with our friends and colleagues. We also want to know what others think is important, current, and of interest, and to show others in our social group that we also are "in the know." The commercial mass media helps to create a common culture that can be shared by many, to some extent even across ethnic and ideological subgroupings.

### C. Trustworthy Supplier of Information

Information has little or no value for either public discourse or individual autonomy if it is inaccurate.[16] Indeed, false and misleading statements of fact can be highly damaging to liberal democratic governance and individual well-being.

---

Charles T. Salmon eds. 1995) [hereinafter PUBLIC OPINION] (noting that given competition among issues for saliency among the public, the public agenda typically consists of no more than five to seven issues).

[14] For further discussion of this point, see Neil Weinstock Netanel, *Asserting Copyright's Democratic Principles in the Global Arena*, 51 VAND. L. REV. 218, 263-67 (1998). See also Owen M. Fiss, *The Censorship of Television*, 93 NW. U.L. REV. 1215, 1217 (1999) (noting that television, unlike today's computer communication, has the capacity to create a shared understanding among a mass audience).

[15] Cass Sunstein and Edna Ullman-Margalit label as "solidarity goods" those "goods whose value increases as the number of people enjoying them increases." Cass Sunstein & Edna Ullman-Margalit, Solidarity in Consumption, John M. Olin Law & Econ. Working Paper No. 98 (2d Series) (April 28, 2000), available at http://papers.ssrn.com/paper.taf?abstract_id=224618.

[16] As the Supreme Court famously noted in Chaplinsky v. New Hampshire, 315 U.S. 568, 571-72 (1942): "There are certain well-defined and narrowly limited classes of speech, the prevention and punishment of which have never been thought to raise any Constitutional problem. These include the lewd and obscene, the profane, the libelous, and the insulting or 'fighting' words – those which by their very utterance inflict injury or tend to incite an immediate breach of the peace.".

Trustworthiness and accuracy in the provision of news and information has long been a central component of journalists' professional standards.[17] Within the mainstream news media, the provision of "a truthful, comprehensive, and intelligent account of the day's events" is seen as the most important responsibility of the press,[18] and the principle of objectivity forms the core of the Code of Ethics of the Society of Professional Journalists.[19] In line with those principles, commercial news media devote considerable resources and professional commitment to checking facts and verifying sources.

*D. The Commercial Media Falls Short of the Fourth Estate Ideal*

As numerous critics have pointed out, the commercial media's actual performance falls short – some would say "far short" -- of the ideal we have just described. In performing its watchdog, discourse-enabling, and information-providing functions, the commercial media skews as well as narrows public debate. Commercial media, critics assert, routinely produce bland, uncontroversial expression, designed to put audiences in a buying mood and to attract a broad cross-section of viewers, readers, and listeners without unduly offending any of them.[20] At the very least, such mainstreaming is unlikely to provide adequate expression to minority interests and concerns. More insidiously, it might help to engender a widespread sense of complacency and a diminished capacity to envision potential challenges to the status quo.[21]

Part of the reason for media's failure to live up to the Fourth Estate Ideal, and one that appears increasingly to be so as media enterprises consolidate into conglomerates with non-media corporate parents, is that media enterprise self-interest and concern for the bottom line pushes coverage to favor commercial interests.[22] Media's skewing also grows out of market dictate, both real and perceived. Sensationalism, mainstream worldview, and reporting that focuses on current political leaders and dominant institutions sells better to broader audiences than alternative content.[23] Moreover, media exhibit considerable herd behavior, imitating the format of existing commercially successful movies, books, and TV shows, thus exacerbating the homogenized, uniform character of much media content.[24] In addition, reporters' dependence on government officials and prominent, well-organized associations for raw material might also vitiate the media's watchdog bite.[25] Some commentators claim, indeed, that government officials exert such a powerful, albeit informal, influence on news coverage and thus on public perceptions and debate that the notion of autonomous media production and distribution of ideas is little more than a pipe dream.[26]

---

[17] U.S. journalists' emphasis on "objectivity" in reporting appears to have had its early beginnings with the penny presses in the 1830s with their self-announced mission to report the "news" and to have attained status as ideology and code of professional conduct after the Progressive Era in the late 19th and early 20th centuries. *See* MICHAEL SCHUDSON, DISCOVERING THE NEWS; A SOCIAL HISTORY OF AMERICAN NEWSPAPERS (1978).

[18] *See* THE COMMISSION ON FREEDOM OF THE PRESS, A FREE AND RESPONSIBLE PRESS: A GENERAL REPORT ON MASS COMMUNICATION: NEWSPAPERS, RADIO, MOTION PICTURES, MAGAZINES, AND BOOKS 21 (1947), quoted in Baker, *supra* note 2, at 348-49.

[19] *See* Society of Professional Journalists, Code of Ethics, available at http://spj.org/ethics/code.htm.

[20] *See, e.g.*, Jerome A. Barron, *Access to the Press: A New First Amendment Right*, 80 HARV. L. REV. 1641, 1641-47 (1967); *see also* James G. Webster & Patricia F. Phalen, THE MASS AUDIENCE: REDISCOVERING THE DOMINANT MODEL 101 (1997) (stating that "(m)any contemporary analysts from both ends of the political spectrum have portrayed the media as inexorably committed to the production of standardized content").

As a result of these factors and others, while the mainstream mass media may often exhibit moderate-reform-oriented norms, it rarely challenges our basic social, economic, and political structures.[27] Nor, for better or for worse, does it provide a full spectrum of fathomable expression and opinion.

---

[21] *See* W. RUSSELL NEUMAN, THE FUTURE OF THE MASS AUDIENCE 28-30 (1991) (summarizing the viewpoint of critical media theorists and other critics that commercial media trivializes political life).

[22] *See* Burt Neuborne, *Media Concentration and Democracy: Commentary*, 1999 ANN. SURV. AM. L. 277, 280.

[23] As Ed Baker demonstrates sensationalist violence, and sex sell particularly well across cultures, and thus assume greater portions of media content as markets for that content become increasingly global. C. Edwin Baker, *International Trade in Media Products, in* THE COMMODIFICATION OF INFORMATION (Niva Elkin-Koren & Neil Netanel eds. forthcoming 2001).

[24] See Neil Weinstock Netanel, *Copyright and a Democratic Civil Society*, 106 Yale L. J. 282, 333-34 (1996); Cass R. Sunstein, *Television and the Public Interest*, 88 Cal. L. Rev. 499, 515-16 (2000).

[25] *See, e.g.*, Clarice N. Olien, et. al., *Conflict, Consensus, and Public Opinion, in* PUBLIC OPINION, *supra* note 13, at 301, 306 (noting media dependency on power relationships); Jonathan Weinberg, *Broadcasting and Speech*, 81 CAL. L. REV. 1103, 1154-55 (1993).

[26] See, e.g., Robert M. Entman, *Putting the First Amendment in Its Place: Enhancing American Democracy Through the Press*, 1993 U. CHI. LEGAL F. 61, 65-72; Steven Shiffrin, *The Politics of the Mass Media and the Free Speech Principle*, 69 IND. L. J. 689, 702-11 (1994).

[27] Weinberg, *supra* note 25, at 1157. At the same time, critics charge that noncommercial, government-subsidized media also generally fall well within the mainstream. See FRANK WEBSTER, THEORIES OF THE INFORMATION SOCIETY 107 (1995) (noting that the BBC's presentation of public affairs has generally "limited itself to the boundaries of established political parties").

### III. Political Party-Financed Media

An alternative model to media that depend upon advertising and sales is media financed in part or in whole by political parties. Party affiliated newspapers dominated the American landscape through much of the first half of the nineteenth-century.[28] The party press also played a major role in post-World War II Western Europe. In recent decades, however, the European party press, like its nineteenth-century American counterpart, has significantly diminished in circulation and influence.[29]

Some commentators lament the passing of an openly ideological, partisan press.[30] They surmise that the politically aligned press brought about considerable political excitement and encouraged higher voting rates and greater citizen participation in politics.[31] Concomitantly, a party press might also sharpen oppositional views and encourage more robust debate.[32] Partisan newspapers in the antebellum United States would not infrequently launch virulent attacks against newspapers of opposing political affiliation,[33] as well as against newspapers, the forerunners of today's commercial press, that charted a more independent course.[34] One might imagine as well that an ideologically driven opposition press would act as a more vigorous watchdog against the government and party in power than does the commercial media.

Yet, the party press system in the antebellum United States ultimately fell into disrepute because the parties in power used government patronage to purchase press loyalty. By the 1820's, newspaper editors regularly received government appointments and subsidies. Newspapers favored by those in power also received lucrative government contracts. These ranged from contracts for printing laws and government documents to concessions for the supply of twine, printed forms, and wrapping paper for the Post Office.[35] Such patronage brought the press into a corrupt bargain. It transformed newspaper editors into "political professionals, people for whom printing was a way to make a living out of politics, rather than the other way around."[36] As John Quincy Adams described the alignment of newspapers behind presidential hopeful and then Adams ally, William Crawford:

> The *National Intelligencer* is secured to him by the belief of the editors that he will be the successful candidate, and by their dependence upon the printing of Congress; the *Richmond Enquirer* because he is a Virginian and slave-holder; the *National Advocate* of New York, through Van Buren, … the *Democratic Press*, of Philadelphia, because I transferred the printing of the laws from that paper to the *Franklin Gazette*; and several other presses in various parts of the Union upon principles alike selfish and sordid.[37]

---

[28] Census data taken in 1850 showed that 95% of all U.S. newspapers had a political affiliation. C. EDWIN BAKER, ADVERTISING AND A DEMOCRATIC PRESS 28 (1994), citing HAZEL DICKEN-GARCIA, JOURNALISTIC STANDARDS IN NINETEENTH-CENTURY AMERICA 48-49, 114-15 (1989).

[29] COUNCIL OF EUROPE PUBLISHING, MEDIA AND ELECTIONS HANDBOOK 11-12 (1999); PETER J. HUMPHREYS, MEDIA AND MEDIA POLICY IN GERMANY; THE PRESS AND BROADCASTING SINCE 1945, 91 (2nd ed. 1994) (noting that in Germany, "[t]he established political press, with its explicit connection to the major parties, has all but disappeared").

[30] *See, e.g.*, BAKER, *supra* note 28, at 41-43.

[31] *Id*. at 41, citing MICHAEL E. MCGERR, THE DECLINE OF POPULAR POLITICS 135 (1986).

[32] See Baker, *supra* note 2, at 358-59 (arguing in support of partisan journalism that "[p]articular groups, especially oppressed groups, … need more segmented or partial dialogues in which to develop their self-conception and their understanding of their own interests").

[33] See Jason P. Isralowitz, Comment, *The Reporter as Citizen: Newspaper Ethics and Constitutional Values*, 141 U. PENN. L. REV. 221, 225, 26 (1992) (describing vituperative debate among early American newspapers, characterized by close ties between editors and party machinery).

[34] *See* SCHUDSON, *supra* note 17, at 54-57 (describing the 1840 "Moral War" of the New York party press establishment against the politically fickle, penny-paper upstart, the *New York Herald*).

Opposition politicians railed against such patronage and its accompanying shackling of the press. They rightly charged that the patronage system transformed the party press into a "government press" and that at the very least, press liberty is compromised when "the favor of power is essential to the support of the editors."[38] But once in power, those critics used the very same tools to reward their loyal supporters and curry favor with newspapers whose future support they sought.[39] It was not until the Civil War, the establishment of the Government Printing Office, and the emergence of a

---

[35] *See* CULVER H. SMITH, THE PRESS, POLITICS, AND PATRONAGE; THE AMERICAN GOVERNMENT'S USE OF NEWSPAPERS 1789-1875 (1977); DONNA LEE DICKERSON, THE COURSE OF TOLERANCE; FREEDOM OF THE PRESS IN NINETEENTH-CENTURY AMERICA 65-71 (1990). The government also provided subsidies for the press on a non-partisan basis. Beginning soon after Independence, Congress heavily subsidized newspaper deliveries by imposing preferential postal rates, levying postal charges on subscribers rather than printers, intermittently collecting subscribers' postal charges, providing free newspaper delivery among printers, and maintaining postal roads for both post office and printers' private use. Richard B. Kielbowicz, *The Press, Post Office, and Flow of News in the Early Republic*, 3 J. EARLY REPUBLIC 255, 257-59 (1983).

[36] Jeffrey Lingan Pasley, "Artful and Designing Men": Political Professionalism in the Early American Republic, 1775-1820, at 336 (Doctoral Thesis, Oct. 1993).

[37] 6 MEMOIRS OF JOHN QUINCY ADAMS 61 (Charles Francis Adams ed. 1874-77; reprinted, Books for Libraries Press, 1969), quoted in DICKERSON, *supra* note 35, at 66.

[38] Duff Green, *U.S. Telegraph*, Feb. 9, 1826, quoted in DICKERSON, *supra* note 35, at 69.

[39] *See* DICKERSON, *supra* note 35, at 65-71 (describing criticism and subsequent use of press patronage by John Quincy Adams and Daniel Webster).

strong independent press that party press patronage as an overt federal government policy came to an end.[40]

To be certain, the party press patronage system in the antebellum United States gained momentum as a result of circumstances unique to that time and place. The high cost of producing and distributing printed material and the absence of a federal government printing office led to the federal government's reliance on the press to print and distribute government documents and the press' financial dependence on government subsidy and patronage. But the antebellum party press' susceptibility to government influence nevertheless presents a cautionary tale. Even short of systematic patronage, there are a multitude of ways in which the government can bestow favors upon media aligned with the party in power, ranging from preferential access to information to subtly discriminatory regulation. As a result, at the same time that the party press aligned with the government will have little incentive to scrutinize government officials and policy, the opposition press will have reduced ability to act as an effective watchdog. In addition, of course, neither government-aligned nor opposition press will have an incentive to question fundamental components of the political system which benefit all major political parties.

Finally, at least at its extreme, a party-aligned press can lead to considerable polarization and insularity. At least within the mainstream of the political spectrum, the commercial press presents an exchange of views even while it supports a common language and understanding for political and social discourse. The party press might have the advantage of challenging mainstream assumptions, empowering marginal voices, and sharpening political debate. But it can also lead to the fragmentation of public discourse, a situation in which debate occurs only with the narrow province of like-minded people. That in turn can lead not only to insularity, but also to greater extremism and polarization.

Cass Sunstein has recently surveyed studies in individual and group psychology documenting the phenomenon of "group polarization."[41] Longstanding democratic theory holds that group deliberation, the exchange of competing views, produces better outcomes. But as Sunstein summarizes, when a group consists of individuals with predeliberation judgments that, on the whole, lean even moderately in a given direction, deliberation tends to move the group, and the individuals who compose it, toward a more extreme position. Thus, with striking empirical regularity, "people who are opposed to the minimum wage are likely, after talking to each other, to be still more opposed; people who tend to support gun control are likely, after discussion, to support gun control with con-siderable enthusiasm; people who believe that global warming is a serious problem are likely, after discussion, to insist on severe measures to prevent global warming; jurors who support a high punitive damages award are likely, after talking, to support an award higher than the median of their predeliberation judgments."[42]

---

[40] *See* SMITH, *supra* note 35, at 229-48.
 41. Cass R. Sunstein, *Deliberative Trouble? Why Groups Go to Extremes*, 110 YALE L.J. 71 (2000).

In short, while our system of free expression might be enhanced by a resurgence of ideologically motivated media, the party press system at best supplement, not replace, independent commercial media. A party press might provide a welcome antidote to the mainstream, market-oriented bias of the commercial media. But it should not be viewed as a panacea.

## IV. Government-Funded Media

Government subsidy provides an opportunity for media to avoid the biases inherent in reliance on advertising and the market for financial sustenance. As such government-funded media can be an important component of the system of free expression. They can serve both to engender a public discourse that is not skewed to the wealthy and provide a forum for minority views that receive little play in commercial media. Indeed, in many democratic countries, state-funded television and other media play important, perhaps even vital, watchdog and discourse-building roles.

However, incidents abound of even democratic governments seeking to use the power of the purse to extract influence over the speech of state-funded media. The problem is not merely the difficulty of setting rational, neutral criteria for the distribution of government subsidies to the press.[43] In addition, state efforts to insulate state-funded broadcasters from government and political party interference have proven to be only partly successful.[44] In some instances, public broadcasters' internal supervisory organs have become politicized along party lines.[45] In others, governments have exerted direct pressure on public broadcasters to alter broadcasting content.[46]

---

42. Cass R. Sunstein, The Law of Group Polarization 1 (U. Chi. Law School, John M. Olin Law & Economics Working Paper No. 91, 1999) available at http://papers.ssrn.com/ paper.taf?abstract_id=199668.

Further, critics charge that noncommercial, government-subsidized media also generally fall well within the mainstream.[47] In part, political pressures constrain public broadcasters from taking controversial positions or even tackling controversial issues.[48] In addition, the people that determine public broadcast programming are largely drawn from the same intellectual and social élites as their counterparts in the commercial media and thus tend to share the same mainstream orientation.

V. Peer-to-Peer Model

Yochai Benkler has argued that a system of distributed, peer-to-peer information production and dissemination would better secure individual autonomy than does our current commercial-media-dominated regime.[49] As Benkler suggests, such a system would be a much-expanded version of non-centralized, peer-to-peer music exchange made possible through Gnutella and other such software.[50] It would entail the establishment of a commons for both content and communications infrastructure. In that way, all could produce and exchange expressive content, ranging from video to text, free from governmental or proprietary control.

---

[43] *See* ANTHONY SMITH, THE POLITICS OF INFORMATION; PROBLEMS OF POLICY IN MODERN MEDIA 174 (1978) (noting that proposals for press subsidies in German have foundered on the difficulty of determining such criteria).
[44] *See* ELI NOAM, TELEVISION IN EUROPE 96-97 (1991) (chronicling decades of post-war French government attempts to influence coverage on state-run French television and radio); *Furore over IRA Film Could Put Peace Talks in Jeopardy*, THE INDEPENDENT (London), July 27, 1997, at 1 (noting that under a United Kingdom government ban, Sinn Fein representatives were not allowed to speak on British television until 1993); *see also* Frances H. Foster, *Information and the Problem of Democracy: The Russian Experience*, 44 AM. J. COMP. L. 243, 257-58 (1996) (detailing instances of government censorship in the supposed "defense of democracy" in post-Soviet Russia); *cf.* FRANCES STONOR SAUNDERS, THE CULTURAL COLD WAR; THE CIA AND THE WORLD OF ARTS AND LETTERS (1999) (documenting the CIA's covert funding of select academic conferences, magazines, and cultural activities in post-war Europe in an effort to lure Western European intelligentsia away from its fascination with Marxism toward a more favorable understanding of the American worldview).
[45] *See* HUMPHREYS, *supra* note 29, at 176-87.
[46] *See* note 44 *supra*.
[47] *See* FRANK WEBSTER, THEORIES OF THE INFORMATION SOCIETY 107 (1995) (noting that the BBC's presentation of public affairs has generally "limited itself to the boundaries of established party politics").
[48] *See* HUMPHREYS, *supra* note 29, at 321 (discussing German public broadcasting).
[49] *See* Yochai Benkler, *Siren Songs and Amish Children: Autonomy, Information, and Law*, 76 N.Y.U. L. REV. 23 (2001).

For Benkler, individual autonomy in the communications environment is best served when individuals enjoy broad opportunities to disseminate expression – both their own creative expression and expression that other's have created – to choose among a highly diverse menu of expression to hear, see, or read. The advantage of his peer-to-peer model is that it takes editorial control away from the commercial firms that currently own carriage media and expressive content. It allows individuals to upload, distribute, and receive the content of their choice.

While Benkler's focus is individual autonomy, a peer-to-peer model might also be seen to better serve democratic institutions by providing for greater expressive diversity and "bottom-up" discourse. Indeed, other commentators have looked to peer-to-peer communication as a vehicle for "democratizing" our system of free expression.[51] As Niva Elkin-Koren puts it: The "transformative power of cyberspace lies in its capability to decentralize the production and dissemination of knowledge".[52]

## VI. Peer-to-Peer; A Critical Evaluation

As I, and others, have argued elsewhere, peer-to-peer communication, such as takes place on the Internet, raises staggering problems of information overload and credibility.[53] Without some vehicle for filtration and accreditation, individuals would face the impossible task of sifting through vast quantities of expression to seek to determine which items are both of interest and reasonably trustworthy. Without tools and institutions for filtration and accreditation, we would be awash in a maelstrom of noise.

---

[50] *Id*. at 107-08.
[51] *See, e.g.,* Dean Colby, *Conceptualizing the "Digital Divide": Closing the "Gap" by Creating a Postmodern Network that Distributes the Productive Power of Speech*, 6 COMM. L. & POL'Y 123 (2001).
[52] Niva Elkin-Koren, *Cyberlaw and Social Change: A Democratic Approach to Copyright Law in Cyberspace*, 14 CARDOZO ARTS & ENT L.J. 215, 217 (1996).
[53] *See* ANDREW L. SHAPIRO, THE CONTROL REVOLUTION; HOW THE INTERNET IS PUTTING INDIVIDUALS IN CHARGE AND CHANGING THE WORLD WE KNOW 133-36, 187-97 (1999); Neil Weinstock Netanel, *Cyberspace 2.0*, 79 TEXAS L. REV. 447, 456 (2000).

Part of the function of the commercial news media is filtration and accreditation. The press often serves as a source of trustworthiness, stability, and accountability, or at least some combination of those traits. Whatever their faults, traditional news media have the resources and professional commitment to check facts and verify sources, and we hold them accountable if they do not. In contrast, Matt Drudge and other individual online publishers have neither the financial wherewithal nor institutional aspiration to meet professional journalistic standards.[54] That is not to say that such gadflies can make no constitutive contribution to public discourse or that the media should hold a monopoly over speech. But full disintermediation of the kind Benkler and others envision would not only Balkanize public discourse; it would also leave us without any real possibility for assessing the reliability and import of vast bulk of expression and opinion swirling around the Internet.

Benkler well recognizes the need for filtration and accreditation in all communication, including peer-to-peer networks.[55] He argues, however, that peer-to-peer networks can generate vehicles for filtration and accreditation that will better serve individual autonomy than filtration and accreditation performed by the commercial news media. First, Benkler argues that filtration and accreditation only enhances autonomy if the editor's notion of relevance and quality comports with those of the sender and recipient and if the editor is herself trustworthy. As he puts it: "To the extent the values of the editor diverge from those of the user, an editor does not facilitate user autonomy by selecting relevant information based on her values, but rather imposes her own preferences regarding what should be relevant to users. A parallel effect occurs with accreditation. An editor might choose to treat as credible a person whose views or manner of presentation draw audiences, rather than necessarily the wisest or best informed commentators."[56] Second, Benkler insists that, armed with the right kind of communications and processing capabilities, peer-to-peer communication can generate its own, bottom-up vehicles for relevance and accreditation.[57] These would include peer-to-peer rating systems, such as are possible with music search engines that track and poll user preferences.[58] In addition to such peer-to-peer rating and referencing, Benkler and others suggest, distributed network communication enable persons and entities to establish competing filtration and accreditation services, giving individuals a broad spectrum of choice regarding which such "authorities" they wish to rely on.[59]

---

[54] SHAPIRO, *supra* note 53, at 133-36.
[55] Benkler, *supra* note 49, at 105-06.
[56] *Id*. at 107.
[57] *Id*. at 107-08.

I will take Benkler's arguments in reverse order, first questioning the efficacy of peer-to-peer filtering and accreditation and then addressing the peer-to-peer model's relative benefits for liberal democratic governance.

### A. The Efficacy of Peer-to-Peer Filtering and Accreditation

Benkler's proffered peer-to-peer filtering and accreditation is of two types: peer-to-peer rating and free lance editing. Peer-to-peer rating might work well with music and other art forms. Music is an "experience good." I can tell what a given musical work is worth to me once I hear it. I might also be able to know from experience that certain persons or certain categories of persons share my musical tastes, such that I can generally rely on their recommendations regarding what music is worth my while to hear.

But the music rating model does not apply to news and other sorts of information. Information is essentially a "credence good."[60] Consumers have difficulty evaluating the value of such a good even after consuming it. Information has value only if it is accurate. And I generally have no means, short of relying on the trustworthiness of the source of information, of knowing whether information I receive about an event or phenomenon outside my personal experience is accurate. (And, of course, if I have such personal knowledge, the report of information has no value to me either because I already know it.)

Peer-to-peer rating over vast communications networks has no mechanism for ensuring trustworthiness. How do I know whether a given piece of "information" is accurate or just rumor, libel, manipulation, or urban legend? Indeed, even music rating is subject to such a problem. I can rely on my peers' music rating only to the extent that the network provides me with accurate information about my peers' preferences. But I have no independent vehicle for judging the accuracy of that information. How do I know whether the ratings reflect true preferences or manipulation by certain groups or musicians, like that which plagues Time Magazine's Person of the Year poll and other such online public opinion polls?[61] (Ratings reflecting music *purchases* might solve this problem, at least to the extent the search engine is trustworthy. But if music must be purchased, we might no longer have a non-proprietary information commons.)

---

[58] *Id.*
[59] *Id.* at 108; *see also* Eugene Volokh, *Cheap Speech and What It Will Do*, 104 YALE L.J. 1805 (1995) (predicting that free lance authors and critics will supplant media firms).
[60] *See* Mark R. Patterson, *On the Impossibility of Information Intermediaries*, Fordham University School of Law, Law and Economics Research Paper No. 13, August 5, 2001, at 5 n. 21, available at http://www.fordham.edu/law/faculty/patterson/workingpapers/infointer.pdf.

Like other situations in which the absence of trustworthiness guarantees results in a paralysis-inducing collective action problem, peer-to-peer network users might turn to trusted institutions to provide credible information and trustworthy assessments of others' information.[62] Such institutions would consist of information intermediaries, like Internet search engines and Consumer Reports, as well competing free lance editors, along the lines of Benkler's model.

But as Mark Patterson has demonstrated, it is highly unlikely that trustworthy information intermediaries would arise to any consistent, systematic extent in the information commons, peer-to-peer environment.[63] Part of the reason stems from the public goods quality of information. Information intermediaries provide information about information produced by others (or aggregate information that the information intermediary warrants as relevant and credible). But once information is released, recipients can pass it on to others at little or no cost. As a result, unless information intermediaries can exert proprietary control over their information, which under Benkler's information commons model they cannot, they will be unable to profit from all

---

[61] The problem of lack of trustworthiness guarantees in peer-to-peer networking has apparently affected music exchange as well. As Sirkka L. Jarvenpaa and Emerson H. Tiller have observed: "Recent research on Gnutella music file sharing, for example, showed that more people take than give - which is the familiar collective action problem. Because giving requires allowing access to one's computer, issues of trust no doubt affect the reciprocal nature of distributed computing and file sharing business arrangements." Sirkka L. Jarvenpaa & Emerson H. Tiller, *Customer Trust In Virtual Environments: A Managerial Perspective*, 81 B.U.L. REV. 665, 667 (2001).

[62] As Jarvenpaa and Tiller note: "Under the rational choice perspective, customers facing prisoner's dilemmas and collective action problems are often willing to trust only to the extent that trust can be fostered by institutions. In such situations, trust takes a form of institution-based trust." *Id.* at 673.

[63] *See, generally,* Patterson, *supra* note 60.

the information they produce and thus will likely produce far less information than socially optimal.[64]

Even worse, information intermediaries have an incentive to profit from slanting the information they provide.[65] In order to maximize their revenue, information intermediaries are likely to sell their services to advertisers and information sources, whether instead of or in addition to selling information to consumers. That has already happened with Internet search engines like Yahoo. Such information intermediaries regularly sell prominent placement in search results.[66] Even so-called "shopping bots," specialized search engines designed to compare prices and goods on behalf of consumers, have been modified to cater to merchants. Shopping bot sites increasingly offer merchants opportunities for partnership, affiliation, advertising, and, most insidiously, top ranking in search results in return for a fee.[67]

To be certain, information intermediaries face certain reputational constraints.[68] But especially where the information concerns the quality and accuracy of news reporting, consumers of information will likely never determine whether the information received was accurate. Given their inability directly to evaluate the quality and credibility of information intermediary services, consumers will need to rely on the same mega-filtration and accreditation tools that apply to the commercial mass media. These involve some combination of information providers' professional ethics, which are already firmly in place in the case of the mass media but which would have to be developed for their peer-to-peer counterparts, and mega-information intermediaries, who report on the accuracy of information intermediaries and who suffer from the same infirmities as any other information intermediary.

Finally, even if commercial, but reasonably credible information intermediaries do emerge, they are unlikely to do so in a competitive market. Like all producers of information, information intermediaries tend to have cost structures characterized by economies of scale. This is especially the case on the Internet, where the marginal cost of information production is extremely low and fixed costs are extremely high. As a result, information industries, online as well as offline, face considerable centripetal force. As economists have long recognized, "where technology creates significant economies of scale, markets tend towards dominance by a few large players."[69] The ongoing consolidations within the media and telecommunications industry are a prime example of this phenomenon. And, as Mark Patterson points out, the market for information intermediaries in the traditional economy also tends strongly towards monopoly.[70] There is no reason to think that peer-to-peer, online networks would somehow spawn a competitive information intermediary market.

---

[64] See id. at 6-7.
[65] See id. at 9-12.
[66] See Lucas D. Introna & Helen Nissenbaum, *Shaping the Web: Why the Politics of Search Engines Matters* 16(3) INFORMATION SOC'Y 169 (2000) (discussing the search engine bias towards firms with resources to gain prominent placement in search results).
[67]. See Karen Solomon, *Revenge of the Bots*, THE STANDARD, Nov. 15, 1999, *available at* http://www.thestandard.com/article/display/1,1151,7624,00.html.
[68] See Patterson, *supra* note 60, at 12-14; Gary Charness & Nuno Garoupa, *Reputation and Honesty in a Market for Information* (Working Paper, Sept. 1998), http://papers.ssrn.com/so13/papers.cfm?abstract_id=139695.

At bottom, the only information intermediaries in the peer-to-peer network world likely to be different in kind from today's mass media are non-profit watchdogs. Such entities already exist in the offline world. Some, such as the Consumers Union, are quite effective in their narrow areas. But non-profit entities are unlikely to assume the role that Benkler envisions in the area of news reporting. Most importantly, the verification of others' reporting requires much the same resources as the reporting itself. And these resources are considerable.[71] Indeed, media critics rightly point out that even the commercial mass media lack the resources to engage consistently in independent investigative reporting, without reliance on government sources to provide information. It is far more costly to track, investigate, and verify ongoing news reporting than to test the safety and reliability of consumer products. While non-profit media watchdogs can sometimes call media news organizations to task for inaccurate and biased reporting, they lack the resources to do so on a consistent and effective basis.

---

[69]. Julie E. Cohen, *Lochner in Cyberspace: The New Economic Orthodoxy of "Rights Management"*, 97 MICH. L. REV. 462, 522 (1998); see also Philip E. Agre, *Life After Cyberspace*, 18 EASST Rev. (Sept. 1999) <http://www.chem.uva.nl/easst/easst993.html>.

[70] *See* Patterson, *supra* note 60, at 4.

[71] *See* PROTESS ET. AL., *supra* note 9, at 233-35 (describing significant investment of news organizations' resources in investigative reporting and in responding to counterattacks by government officials and private firms who are subjects of exposés).

### B. *Peer-to-Peer's Relative Benefits for Liberal Democratic Governance*

Even if peer-to-peer networks could generate information producers and intermediaries along the lines Benkler suggests, we would not want to jettison the commercial mass media. Benkler, it seems, imagines a two-dimensional speech universe populated entirely by volunteer individual speakers revolving around a benign government.[72] To my mind, however, liberal democratic governance requires, rather, a speech universe punctuated by bubbles, a universe that contains concentrations of private expressive power capable of standing up to government and, indeed, capable of standing up to General Motors, Mobil Oil, the Republican and Democratic parties, the World Bank, and other concentrations of private economic and political power. Democratic governance requires a free press not just in the sense of a diversity of expression. It requires the *institution* of a free press. It requires media with the financial wherewithal and political independence to engage in sustained investigative journalism, to expose the errors and excesses of government and other powerful political and economic actors.

The Benklerian republic of yeoman speakers presupposes a republic of yeomen, period. It imagines a world in which we earn our livings, communicate, and govern ourselves without mediating organizations and concentrations of wealth and power. But that is not our world, never has been our world, and, so long as we are organized in polities larger than the Greek city-state, never will be our world. In fact, I dare say, it is not a world to which we should aspire because in such a world, nothing would stand between the individual and an all-powerful state.

Our best hope for democratic governance in this world is far messier than the ideal republic of yeomen. It requires mediating institutions and associations, private and public concentrations of wealth and power, and varied mechanisms to maintain multiple balances of power within government, within civil society, and between government and civil society.[73] No less, it requires an expressive sector that mirrors that panoply of governing institutions. It requires independent, powerful media to guard against abuses of power on the part of other private and governmental entities. It also requires mechanisms to check and distribute the power of the commercial media.

---

[72] *See* Yochai Benkler, *Free as the Air to Common Use: First Amendment Constraints on Enclosure of the Public Domain*, 74 N.Y.U. L. REV. 354, 400 (1999) (complaining that an expanded copyright "tends to produce market-based hierarchy, rather than to facilitate and sustain independent yeoman authors").
[73] Intermediate associations can also help to support a democratic culture by building participatory norms. See Grant McConnell, *The Public Values of the Private Association, in* NOMOS XI: VOLUNTARY

Individual authors and web site operators lack the resources to fulfill the press' traditional, vital role of watchdog against government myopia and oppression. Nor can individuals adequately expose corporate unlawfulness, labor union corruption, and political party self-aggrandizement. Liberal democratic nations necessarily encompass multiple concentrations of power.[74] Only an equally powerful press can effectively check other entities' and associations' deployment of their power by exposing it to the light of public opinion. Indeed, only a *mass* media, capable of reaching a mass audience, can both catalyze and, to a degree, embody public opinion in the face of government authority and corporate fiefdom.

The story of former tobacco company researcher Jeffrey Wigand is a case in point.[75] While employed at cigarette-maker Brown & Williamson, Wigand discovered that, despite repeated public denials and testimony under oath to the contrary, leading tobacco company executives had long known that the nicotine in cigarettes is an addictive substance. In 1995 Wigand relayed his discovery to a reporter for the widely watched CBS news magazine, 60 Minutes, and agreed to be interviewed on the program. In so doing, Wigand violated his Brown & Williamson non-disclosure agreement and undertook considerable financial and personal risk. Instrumental in Wigand's decision to divulge the tobacco company misdeeds despite that risk was CBS's promise to indemnify him for any liability to his former employer and, no less importantly, the knowledge that he would have the opportunity to present his findings before millions of prime time television viewers.[76] It is highly unlikely that Wigand would have exposed Brown &

---

ASSOCIATIONS 147, 149 (J. Pennock & J. Chapman eds. 1969). My focus here, however, is on intermediate organizations as loci of power vis-à-vis other power centers.

[74] *See* PETER H. SCHUCK, THE LIMITS OF LAW; ESSAYS ON DEMOCRATIC GOVERNANCE 204-50 (2000). *See also* Mark P. Petracca, *The Future of an Interest Group Society*, in THE POLITICS OF INTERESTS; INTEREST GROUPS TRANSFORMED 345 (Mark P. Petracca ed. 1992) (noting the continuing, vital role played by interest groups in the democratic process).

[75] My account of the Wigand story draws heavily upon Paul Starr, *What You Need to Beat Goliath*, AM. PROSPECT, Dec. 20, 1999, at 7.

[76] Some ten million households watch 60 Minutes each week. Owen M. Fiss, *The Censorship of Television*, 93 NW. U.L. REV. 1215, 1217 (1999) (citing Nielsen Media Research, 1998 Report on Television (1999)).

Williamson at his own peril without the backing and mass audience of a major media outlet. And even if he had, perhaps by posting information and documents on his personal web page, his story may well have been lost in the chorus of tobacco company denials (if they had even bothered to respond) and against the backdrop of tens of thousands of crank web pages presenting asundry allegations that few find credible even when true.

Of course, there is more to the Wigand story. CBS management initially scuttled the Wigand interview shortly before it was to appear on 60 Minutes. Apparently, CBS did want to take the unlikely, but not immaterial risk of having to pay a multi-billion dollar damage award to Brown & Williamson, especially since that contingency would have reduced CBS's share price at a time when the company was negotiating to be acquired by Westinghouse. Additional, and more insidious, corporate entanglements might have also contributed to the decision.[77] Laurence Tisch, CBS's chairman at the time, was also an owner of Lorillard Tobacco Company, and his son Andrew was Lorillard's president. In fact, Andrew Tisch was one of the tobacco executives who had sworn before Congress that nicotine was not addictive, and Wigand was a witness in the perjury investigation regarding that testimony.[78] At that time, moreover, Lorillard was negotiating with Brown & Williamson to buy six of its brands.[79] In sum, CBS's broadcast of the Wigand interview might have caused CBS and Laurence Tisch considerable financial loss and might have helped send Tisch's son to jail, facts that could hardly have been lost on the CBS lawyers and executives involved in the decision to cancel the Wigand broadcast.[80]

The circumstances surrounding CBS's cancellation of the Wigand interview graphically illustrate the potential vulnerabilities and limitations of the media-as-watchdog model, especially in an age of increasing media consolidation and conglomeration. But some three months after the Wigand interview was to have been aired, CBS reversed its decision to scuttle the interview and did in fact broadcast it; and in so doing, it dealt a significant blow to the tobacco industry. Significantly, it was the presence and coverage of competing media that pushed a reluctant CBS into reasserting its watchdog role. CBS broadcast the Wigand interview only after the New York Times had detailed CBS's capitulation in a front page story,[81] the New York Daily News had obtained and reported on the transcript of the omitted interview,[82] and the Wall Street Journal had published Wigand deposition testimony containing his central allegations.[83] Moreover, critical media coverage did not end with the CBS broadcast. Most notably perhaps, in its 1999 motion picture, The Insider, Disney/ABC presented a widely-acclaimed dramatized version of the entire Wigand episode, castigating CBS for the network's striking, if temporary, abdication of its journalistic integrity. That contemporaneous and subsequent coverage by competing media enterprises, an instance of such enterprises exposing each other's wrongdoing, may deter such lapses in the future.

---

[77] Starr, *supra* note 75.
[78] Id.
[79] Id.
[80] Id.

The moral I wish to draw from this story is not that "all's well that ends well," that we can complacently rely on an increasingly conglomeratized mass media to expose corporate wrongdoing. Rather, the Wigand story illustrates that, like a host of governmental and private institutions, media power has both a positive and negative face. There are no perfect solutions to this dual face of power. There are only partial solutions, most involving efforts to balance one power center against one or more others. State subsidies for non-commercial media, state-imposed limitations on media firm conglomeration, and state assurance of individual access to effective channels of communication constitute such partial solutions. These measures aim to provide for greater expressive diversity without undermining the media's constitutive force. They are not perfect solutions. They leave considerable market hierarchy in place, rendering us vulnerable to media co-option and market suppression of minority voices. But such measures, and others like them, are the best that we can or should hope to obtain. To wipe out the power of the commercial media would simply throw out the proverbial baby with the bath water.

---

[81] Bill Carter , "*60 Minutes" Ordered to Pull Interview in Tobacco Report*, N.Y. TIMES, at A1 (Nov. 9, 1995).
[82] Joe Calderone and Kevin Flynn, *What "60 Minutes" Cut; Attack on Cig Maker Was Axed, Transcript Shows*, N.Y. DAILY NEWS, at 2 (Nov. 17, 1995).

Conclusion

Our system of expression should include considerable space for decentralized, individual expression, such as on the Internet. In addition to constituting a vast forum for alternative discourses, the Internet spawns gadflies, media critics and others may sometimes successfully bring a story to mass media attention or challenge media silence.[84] But that is not to say that we should aspire, even as a liberal democratic ideal, to an egalitarian expressive universe composed entirely of virtual and street-corner pamphleteers. Rather, our system of free expression must include a plurality of speaker types, including commercial mass media, government-subsidized noncommercial media, independent publishers, political and nonprofit associations, universities, and individuals. To some extent, each of these speaker types offsets, complements, and checks the rest. Like other speaker types, the commercial media plays a unique and vital role within this complex mix.

---

[83] See Bill Carter, *CBS Broadcasts Interview With Tobacco Executive*, N.Y. TIMES, at B8 (Feb. 5, 1996) (reporting that CBS broadcast the interview a week after the Wall Street Journal had published the transcript of the Wigand deposition).
[84] An already classic example is Matt Drudge's web site scoop of the Monica Lewinsky story. *See* SHAPIRO, *supra* note 53, at 41.